\documentclass[prl,twocolumn,showpacs]{revtex4}
\usepackage{graphicx}

\begin{document}
\title{Variation of Superconducting Transition Temperature in Hole-Doped Copper-Oxides }
\author{X. J. Chen }
\affiliation{ Department of Physics, Kent State University, Kent, Ohio 44242 }
\author{H. Q. Lin }
\affiliation{ Department of Physics, The Chinese University of Hong Kong, Hong Kong, China }
\date{\today}

\begin{abstract}
The experimentally observed difference of superconducting critical temperature $T_{c}$ in 
hole-doped cuprates is studied by using an extended interlayer coupling model for layered 
$d$-wave superconductors. We show that the change of the maximum $T_{c}$ from series to 
series is determined by the next nearest neighboring hopping $t^{\prime}$, while the 
difference of the maximum $T_{c}$ among the compounds in a homogeneous series is controlled 
by the interlayer pairing strength. Our results provide helpful guidelines in the search for
new high-$T_{c}$ superconductors.      
\end{abstract}
\maketitle

The nature of high temperature superconductors is a challenge problem in condensed matter 
physics. A common feature of copper oxide superconductors is the presence of CuO$_{2}$ plane. 
It has been observed that the superconducting critical temperature $T_{c}$ varies parabolically
with the hole concentration $n_{H}$ in CuO$_{2}$ plane with a maximum $T_{c}^{max}$ at an 
optimal doping level \cite{torr,ando}. However, $T_{c}^{max}$ attainable is different from 
series to series, $e.g.$ $35$ K in La$_{2-x}$Sr$_{x}$CuO$_{4}$ \cite{cava} and $97$ K in 
HgBa$_{2}$CuO$_{4+\delta}$ \cite{wagn}. An obvious question is what is the crucial parameter 
that governs the $T_{c}^{max}$ of each family.

Among various parameters proposed, the Madelung potential at the apical oxygen relative to 
that at the planar oxygens \cite{ohta} was found to correlate with $T_{c}^{max}$ rather 
well, pointing to the primary importance of the apical oxygens for the electronic structure 
relevant to superconductivity. Further investigations \cite{raim,pava} revealed that the 
effect of the apical oxygens on high-$T_{c}$ superconductivity in reality translates into a 
correlation between $T_{c}^{max}$ and the next nearest neighbor hopping parameter 
$t^{\prime}$ in the $t$-$t^{\prime}$-$J$ model. In these approaches, $t^{\prime}$ was 
considered as a single parameter reflecting the main difference among various cuprates.
If we consider the homologous series, the universality of such a correlation would be 
seriously questioned. For example, the bilayer and trilayer Tl$_{2}$-based and Hg-based 
compounds have almost same $t^{\prime}$ \cite{pava}, but their $T_{c}^{max}$'s are 
significantly different. 

Our goal in this work is to extract and identify which parameters govern the $T_{c}$ 
behaviors in hole-doped cuprates. We apply an interlayer coupling model to CuO$_{2}$ layer 
systems and then calculate $T_{c}$ based on the BCS gap equation with $d$-wave symmetry.
Our results suggest that the difference of $T_{c}^{max}$ from series to series is the   
result of different next nearest neighbor hopping $t^{\prime}$, while the difference of 
$T_{c}^{max}$ between the compounds in a homologous series is controlled by the interlayer 
coupling strength $T_{J}$.

The effective layered Hamiltonian we consider is 
\begin{eqnarray}
H&=&\sum_{lk\sigma}\left(\varepsilon _{k}-\mu \right)c_{k\sigma}^{\dagger l}%
c_{k\sigma }^{l}
- \sum_{lkk^{\prime }}V_{kk^{\prime }}c_{k{\small \uparrow}}^{\dagger l}%
c_{-k{\small \downarrow }}^{\dagger l}c_{-k^{\prime }{\small \downarrow }}^{l}%
c_{k^{\prime }{\small \uparrow }}^{l} \nonumber \\
&+& \sum_{<ll^{\prime }>}\sum_{k}T_{J}(k)c_{k{\small \uparrow }}^{\dagger l}%
c_{-k{\small \downarrow }}^{\dagger l}c_{-k{\small \downarrow }}^{l^{\prime }}%
c_{k{\small \uparrow }}^{l^{\prime }}~~,
\end{eqnarray}
where $\varepsilon _{k}$ is the quasiparticle dispersion, $\mu$ is the chemical potential, 
$c_{k{\small \sigma}}^{\dagger l}$ is a quasiparticle creation operator pertaining to the 
layer ($l$) with two-dimensional wave-vector $k$ and spin ${\small \sigma}$. The summation 
over $ll^{\prime }$ runs over the layer indices of the unit cell. The intralayer interaction 
$V_{kk^{\prime }}$ is assumed to be independent of $l$. The interlayer tunneling is 
parameterized by $T_{J}(k)=T_{J}(\cos k_{x}-\cos k_{y})^{4}$ \cite{chak}.

We assume that the superconducting gap is characterized by the nonvanishing order parameter
$b_{k}^{l}=<c_{k{\small \uparrow }}^{l}c_{-k{\small \downarrow }}^{l}>$. Based on the BCS 
theory the gap function $\Delta _{k}^{l}$ satisfies the following equation
\begin{equation}
\Delta _{k}^{l}=\sum_{k^{\prime }}V_{kk^{\prime }}b_{k}^{l} + 
T_{J}(k)(b_{k}^{l+1}+b_{k}^{l-1})~~,
\end{equation}
where $b_{k}^{l}=\Delta _{k}^{l}\chi _{k}^{l}$ and the generalized pair susceptibility is
$\chi _{k}^{l}=(2E_{k}^{l})^{-1}\tanh(\beta E_{k}^{l}/2)$. The quasiparticle spectrum is 
$E_{k}^{l}=\sqrt{(\varepsilon_{k}-\mu)^{2}+\left|\Delta_{k}^{l}\right|^{2}}$.

The spatial dependence of the gap is taken the form \cite{bycz}: $\Delta _{k}^{l}=\Delta 
_{k}^{\pm}e^{\pm i\alpha l}$. Then the general solution of the homogeneous part is
\begin{eqnarray}
\Delta _{k}^{l}=\Delta  _{k}^{+}e^{i\alpha l}+\Delta  _{k}^{-}e^{-i\alpha l}~~.
\end{eqnarray}
Considering the fact that the gap vanishes on the layer ends $l=0$ and $n+1$, the natural 
boundary conditions for the gap are $\Delta_{k}^{0} = \Delta_{k}^{n+1}\equiv 0$. The wave 
vector of the oscillating gap is then determined by  
\begin{eqnarray*}
\left( \begin{array}{clcr}
1 & 1\\
e^{i\alpha l} & e^{-i\alpha l}
\end{array} \right) \left( \begin{array}{clcr}
\Delta  _{k}^{+} \nonumber \\ \Delta  _{k}^{-}
\end{array} \right)=0~~. \nonumber \\
\end{eqnarray*}
The vanishing determinant of the matrix provides a nontrivial solution only when $\alpha = 
\xi \pi /(n+1)$ where $\xi $ is an integer. Thus we obtain $\Delta _{k}^{+}=-\Delta  
_{k}^{-}\equiv \Delta  _{k}$. The solution of spatial dependence of the gap is then
\begin{eqnarray}
\Delta _{k}^{l}=2i\Delta _{k} \sin \left(\frac{\xi \pi m}{n+1}\right)~~.
\end{eqnarray}
The solution with the lowest energy is nodeless inside the $n$ CuO$_{2}$ layers which leads 
to $m=1$ for the superconducting state.

\begin{figure}[tbp]
\begin{center}
\includegraphics[width=\columnwidth]{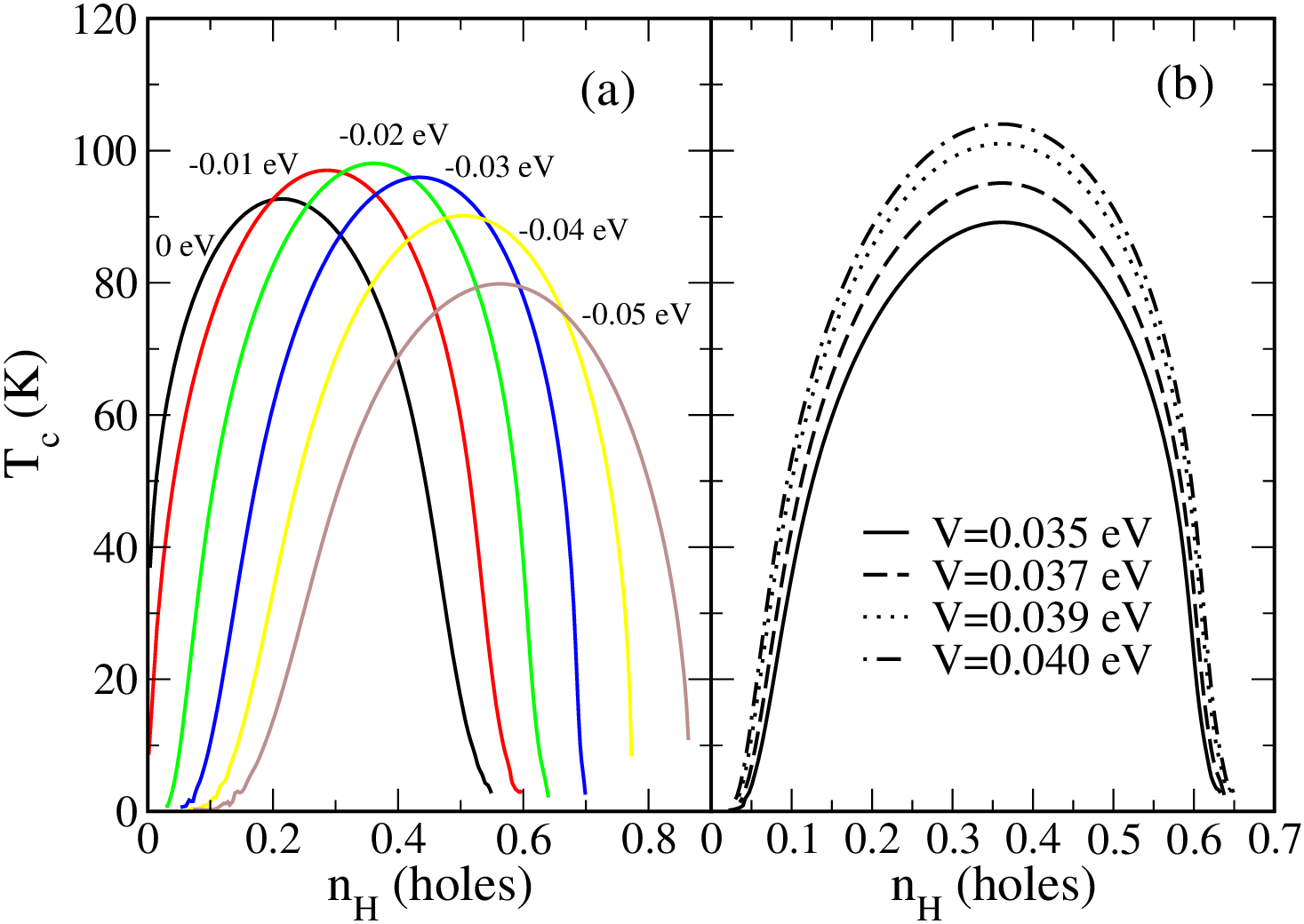}
\end{center}
\caption{ $T_{c}$ vs $n_{H}$ for various $t^{\prime}$ with $V=$0.038 eV (a) and for 
various $V$ with $t^{\prime}=-0.02$ eV (b) in monolayer cuprates. }
\end{figure}

Around critical temperature $T_{c}$, we can take $\chi _{k}^{l}$ in a simple form:
$\chi _{k}^{l}\simeq (2E_{k})^{-1}\tanh(\beta E_{k}/2)\equiv \chi _{k}$ with
$E_{k}=\sqrt{(\varepsilon _{k}-\mu)^{2}+\left|\Delta_{k}\right|^{2}}$. Then gap magnitude 
$\Delta _{k}$ can be written as
\begin{eqnarray}
\Delta _{k}-\sum_{k^{\prime }}V_{kk^{\prime }}\chi _{k^{\prime }
}\Delta_{k^{\prime }}=f(n)T_{J}(k)\chi _{k}\Delta_{k}~~,
\end{eqnarray}
where $f(n)=2\cos (\pi /(n+1))$.

To account for the experimental observed $d$-wave gap, we assume a $d$-wave pairing potential
\begin{eqnarray}
V_{kk^{\prime }}=Vg(k)g(k^{\prime })~~,~~ g(k)=\cos k_x - \cos k_y ~~,
\end{eqnarray}
The gap magnitude is thus $\Delta _{k}=\Delta _{0}g(k)$ and the parameter $\Delta _{0} $ is 
determined by the following self-consistent equation:
\begin{eqnarray}
1=\frac{1}{2N}\sum _{k}\frac{Vg^{2}(k)+f(n)T_{J}(k)}{E_k}\tanh(\frac{\beta E_k}{2})~~.
\end{eqnarray}
The value of $T_{c}$ in layered $d$-wave superconductors is then obtained by solving Eq. (7) 
at $\Delta _{0} =$0.

\begin{figure}[tbp]
\begin{center}
\includegraphics[width=\columnwidth]{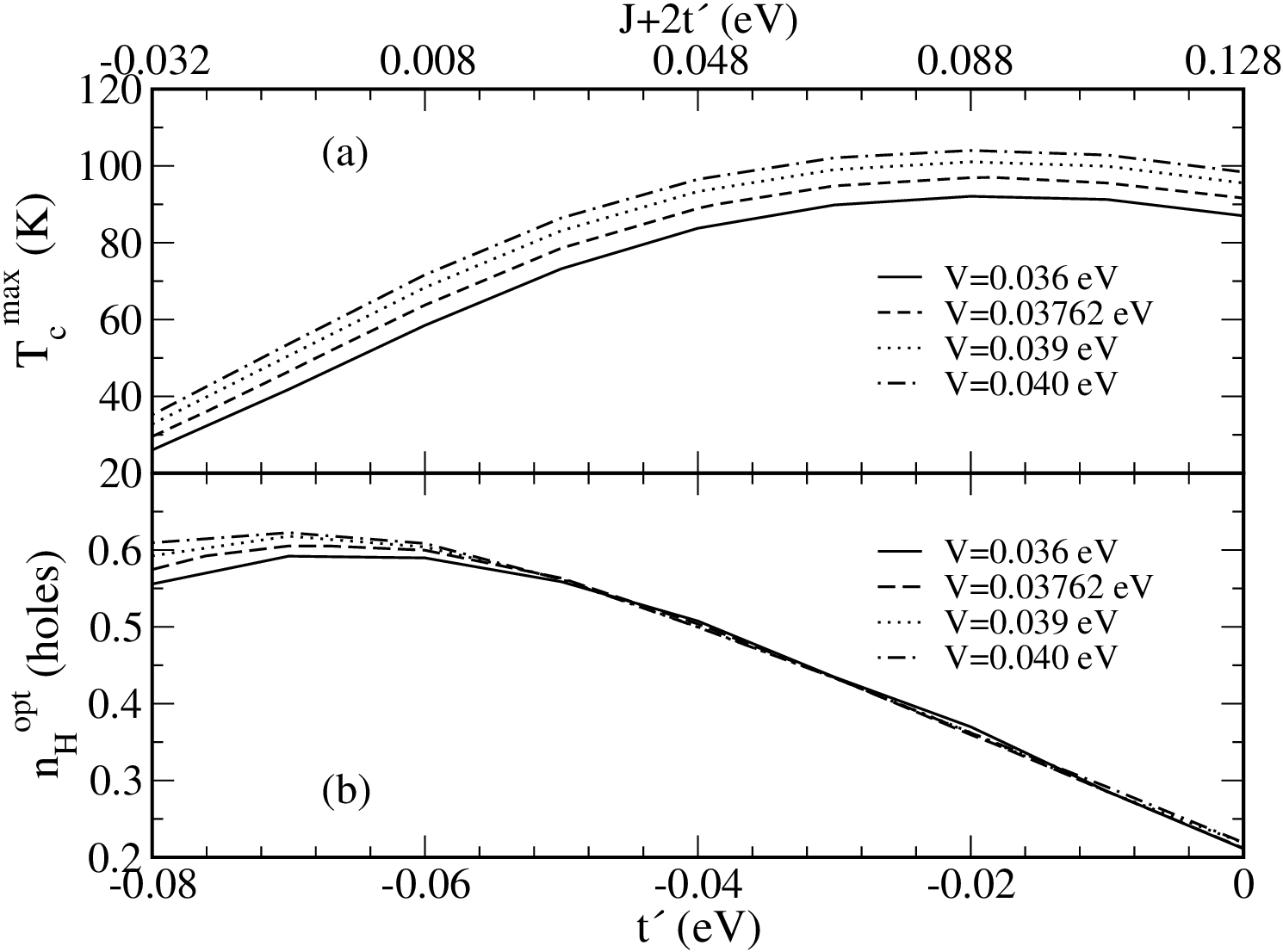}
\end{center}
\caption{ Calculated $T_{c}^{max}$ (a) and $n_{H}^{opt}$ (b) as a function of $t^{\prime}$ 
(or $J+2t^{\prime}$) for various $V$ in monolayer cuprates. }
\end{figure}

In order to self-consistently calculate $T_{c}$ for a given $\mu$ in conjunction with the 
equation determining $n_{H}$, we need an explicit form of $\varepsilon_{k}$. It has been 
established \cite{hole} that the quasiparticle excitation spectrum of cuprates can be well 
described by the $t$-$t^{\prime}$-$J$ model. Within the framework of the $t$-$t^{\prime}$-$J$ 
model, the dispersion $\varepsilon _{k}$ is given by 
\cite{maek,beli} 
\begin{equation}
\varepsilon _{k}= (J+2t^{\prime})\cos k_{x}cos k_{y}
         + \frac{J}{4}(\cos 2k_{x}+cos 2k_{y})~~.
\end{equation} 
For monolayer insulator La$_{2}$CuO$_{4}$, experiments \cite{sing} and theoretical 
calculations \cite{hybe} give a $J=$0.128 eV. There are small variations of $J$ among various 
Cu-O insulators \cite{sule} but we expect a value of $J=$0.128 eV is a generally good 
representation for all Cu-O materials. Then one can determine $T_{c}$ as a function of
$n_{H}$ based on Eqs. (7) and (8) once having knowledge of $t^{\prime}$, $V$, or/and $T_{J}$.

\begin{table*}
\caption{\label{tab:table} Summary of the experimental results of the critical temperature 
$T_{c}^{max}$ at optimal doping, the distance $d_{Cu-O(a)}$ between the copper and apical 
oxygen atoms, the distance $d_{Cu-O(p)}$ between the copper and in-plane oxygen atoms, and 
the calculated values of the bond valence sums of copper $V_{Cu}$ and the difference in the 
Madelung site potentials $\Delta V_{M}$ for a hole between the in-plane oxygen and copper 
atoms in some typical monolayer cuprates. }
\begin{ruledtabular}
\begin{tabular}{cccccc}
Cuprates & $T_{c}^{max}$ (K) & $d_{Cu-O(a)}$ ($\AA $) & $d_{Cu-O(p)}$ 
($\AA $) & V$_{Cu}$ & $\Delta V_{M}$ (eV) \\
\hline
La$_{1.85}$Sr$_{0.15}$CuO$_{4}$ & 35 & 2.4124 & 1.8896 & 2.539 & 49.620 \\
Bi$_{2}$Sr$_{1.61}$La$_{0.39}$CuO$_{6+\delta}$ & 36 & 2.461 & 1.901 & 2.437 & 48.437 \\
TlBa$_{1.2}$La$_{1.8}$CuO$_{5}$ & 52 & 2.500 & 1.9240 & 2.280 & 48.409 \\
Tl$_{2}$Ba$_{2}$CuO$_{6}$ & 90 & 2.714 & 1.9330 & 2.135 & 47.081 \\
HgBa$_{2}$CuO$_{4+\delta}$ & 97 & 2.780 & 1.9375 & 2.091 & 46.81 \\
\end{tabular}
\end{ruledtabular}
\end{table*}

\begin{table*}
\caption{\label{tab:table} The critical temperature $T_{c}^{max}$ and the ratio of $T_{J}/V$ 
in homogeneous copper-oxides series at optimal doping. The brackets are the experimental data
taken from the works of Refs. \cite{ando,cava,wagn,ohta,subr,tora,aiyo,chen1,chen2}. }
\begin{ruledtabular}
\begin{tabular}{cccccccc}
$n$ & 1 & 2 & 3 & 4 & 5 & $\infty$ & $T_{J}/V$ \\
\hline
Bi$_{2}$Sr$_{2}$Ca$_{n-1}$Cu$_{n}$O$_{2n+4+\delta}$ & 36.0 (36) & 90.0 (90)& 115.5 (110) & 
127.8 & 134.7 & 150.7 & 0.1945 \\
Tl$_{2}$Ba$_{2}$Ca$_{n-1}$Cu$_{n}$O$_{2n+4+\delta}$ & 90.0 (90) & 115.0 (115) & 125.2 (125) & 
130.1 (116) & 132.9 & 139.4 & 0.0906 \\
TlBa$_{2}$Ca$_{n-1}$Cu$_{n}$O$_{2n+3+\delta}$ & 52.0 (52) & 107.0 (107) & 131.3 (133.5) & 
143.0 (127) & 149.5 & 164.6 & 0.1930 \\
HgBa$_{2}$Ca$_{n-1}$Cu$_{n}$O$_{2n+2+\delta}$ & 97.0 (97) & 127.0 (127) & 139.2 (135) & 145.2 
(129) & 148.6 (110) & 156.4 & 0.1135 \\
\end{tabular}
\end{ruledtabular}
\end{table*} 

First we consider the variation of $T_{c}$ in monolayer ($n=1$) hole-doped cuprates. Figures 
1 (a) and (b) show the calculated $T_{c}$ in monolayer superconductors as a function of 
$n_{H}$ in some interested parameters range of $t^{\prime}$ and $V$. As shown, $T_{c}$ 
initially increases with increasing $n_{H}$, takes a maximum around an optimal doping level 
$n_{H}^{opt}$ and then decreases with further increasing $n_{H}$. This parabolic relation 
between $T_{c}$ and $n_{H}$ agrees with general experimental observations in monolayer 
cuprates \cite{torr,ando}. We notice that $T_{c}^{max}$ systematically changes with 
$t^{\prime}$, but it monotonically increases with $V$, as one expects. The difference between 
these two parameters is that $n_{H}^{opt}$ depends significantly on $t^{\prime}$, while it 
scarcely changes for different values of $V$. These results indicate that the parameters 
controlling $T_{c}^{max}$ would be either $t^{\prime}$ or $V$ or both of them.

In Fig. 2 we plotted the $t^{\prime}$ dependence of both $T_{c}^{max}$ and $n_{H}^{opt}$ for 
monolayer cuprates. As $t^{\prime}$ increases, $T_{c}^{max}$ increases and then decreases 
through a maximum for all $V$ studied. The occurrence of the maximum implies that the 
enhancement of $T_{c}^{max}$ due to the increase in $t^{\prime}$ is limited. $n_{H}^{opt}$ 
behaves in a similar manner with $t^{\prime}$ as $T_{c}^{max}$. For $J+2t^{\prime}>0$, 
$n_{H}^{opt}$ decreases with increasing $t^{\prime}$. Although $T_{c}^{max}$ depends on $V$, 
$n_{H}^{opt}$ is nearly independent of $V$ over a wide range of $t^{\prime}$.

To trace the clue to the change of $T_{c}^{max}$ among monolayer cuprates, we list in Table I 
the experimental results of $T_{c}^{max}$ \cite{cava,ando,subr,tora,wagn}, the distance 
$d_{Cu-O(a)}$ between the copper and apical oxygen atoms and the distance $d_{Cu-O(p)}$ 
between the copper and in-plane oxygen atoms taken from the works in Refs. \cite{wagn,ohta},
the calculated values of bond valence sums (BVS) of copper $V_{Cu}$ and the difference in the 
Medelung site potential for a hole between the copper and the in-plane oxygen $\Delta V_{M}$. 
To get effective BVS of copper, we follow the method proposed by Brown \cite{brown}. The 
results of $\Delta V_{M}$ based on the structural data are taken from the works in Refs. 
\cite{ohta,muroi}. Here we observe one important experimental fact: $T_{c}^{max}$ increases 
systematically with enlarging $d_{Cu-O(a)}$. Band structure calculations \cite{pava} revealed 
that $t^{\prime}$ increases with $d_{Cu-O(a)}$ for the monolayer cuprates reported so far.
Thus the increase of $T_{c}^{max}$ with increasing $t^{\prime}$ should capture the basic 
physics of the monolayer cuprates.

It has been proposed \cite{leeu,tana} that $V_{Cu}$ and $\Delta V_{M}$ are two essential 
factors governing $T_{c}$ and represent an essentially equivalent physical content. Materials 
with larger $T_{c}^{max}$ tend to have a smaller $V_{Cu}$ \cite{leeu} or $\Delta V_{M}$ 
\cite{tana}. Since the variation of $V_{Cu}$ or $\Delta V_{M}$ reflects the corresponding 
change of $n_{H}$ \cite{chen1,tana,whan}, the increase of the calculated $T_{c}^{max}$ with 
decreasing $n_{H}^{opt}$ for a wide $t^{\prime}$ range is obviously consistent with the 
experimental data shown in Table I. This $n_{H}^{opt}$ dependence of $T_{c}^{max}$ is also 
consistent with the muon spin resonance ($\mu$SR) measurements \cite{uemu}. On the other 
hand, the fact that the change of $T_{c}^{max}$ with $V$ is almost independent of 
$n_{H}^{opt}$ (Fig. 2(b)) rules out the possibility of $V$ being a dominant factor in 
governing the change in $T_{c}^{max}$. The present results lead us to conclude that 
the increase of $T_{c}^{max}$ with $d_{Cu-O(a)}$ among the monolayer cuprates is a result of 
the increase in $t^{\prime}$. One prediction is that $T_{c}^{max}$ decreases with further 
increasing $t^{\prime}$ after through a saturation. Thus, materials with a relatively long 
$d_{Cu-O(a)}$ bondlength would not expect to have a high $T_{c}^{max}$.  

The values of $t^{\prime}$ were determined in a self-consistent way as follows. From Fig. 2 
(a) we learned that there exists a maximum for given $V$. Among the monolayer cuprates 
discovered so far, HgBa$_{2}$CuO$_{4+\delta}$ possesses the highest $T_{c}^{max}$ of 97 K.
Assuming this is the highest value in all monolayer cuprates, we derived a value of 
$V=0.03762$eV from curves of $T_{c}^{max}$ versus $t^{\prime}$. Equation (7) yields 
$t^{\prime}=-0.0183$ eV for the optimally doped HgBa$_{2}$CuO$_{4+\delta}$. For other 
optimally doped monolayer compounds with $T_{c}^{max}<97$K, $t^{\prime}$ should be smaller 
than $-0.0183$ eV because of their shorter $d_{Cu-O(a)}$. The relative $t^{\prime}$ is then 
obtained by using the experimentally observed $T_{c}^{max}$.     

Next we consider $n$, the number of CuO$_{2}$ layers, dependence of $T_{c}$ in the layered 
homogeneous series. In general, $T_{c}^{max}$ initially increases with $n$, maximizes at 
$n=3$, and then decreases with further increasing $n$ \cite{aiyo}. To calculate $T_{c}$ for 
multilayers, we use the same dispersion $\varepsilon_{k}$ and $V$ as obtained from the 
monolayer. The interlayer tunneling strength $T_{J}$ is determined by using the experimental 
values of $T_{c}^{max}$ for monolayer and bilayer compounds in the same homogeneous series.
As an example, in Fig. 3, we show curves of calculated $T_{c}$ versus $n_{H}$ as a function 
of layer number $n$ in the Hg-based series. The parabolic behavior is generally observed for 
any layered compound. The calculated $T_{c}^{max}$ in four typical homogeneous series are 
summarized in Table II. The experimental results are also listed for comparison. As can be 
seen, $T_{c}^{max}$ initially increases with increasing $n$ and then saturates as $n 
\rightarrow \infty$. This behavior is in good agreement with those obtained from both the 
interlayer mechanism \cite{bycz,whea} and Ginzburg-Landau theory \cite{birm,chen2}. The 
upper limit of $T_{c}^{max}$ for infinite layer compound is in the range of 139.4 to 164.6 K. 
The highest $T_{c}^{max}$ of 164.6 K is found in the Tl-based series. Our results for $n=3$ 
agree with experiments very well. The predictions made here for $T_{c}^{max}$ of the trilayer 
compound is the best ones compared to previous theories \cite{bycz,whea,birm,chen2}. 

\begin{figure}[tbp]
\begin{center}
\includegraphics[width=\columnwidth]{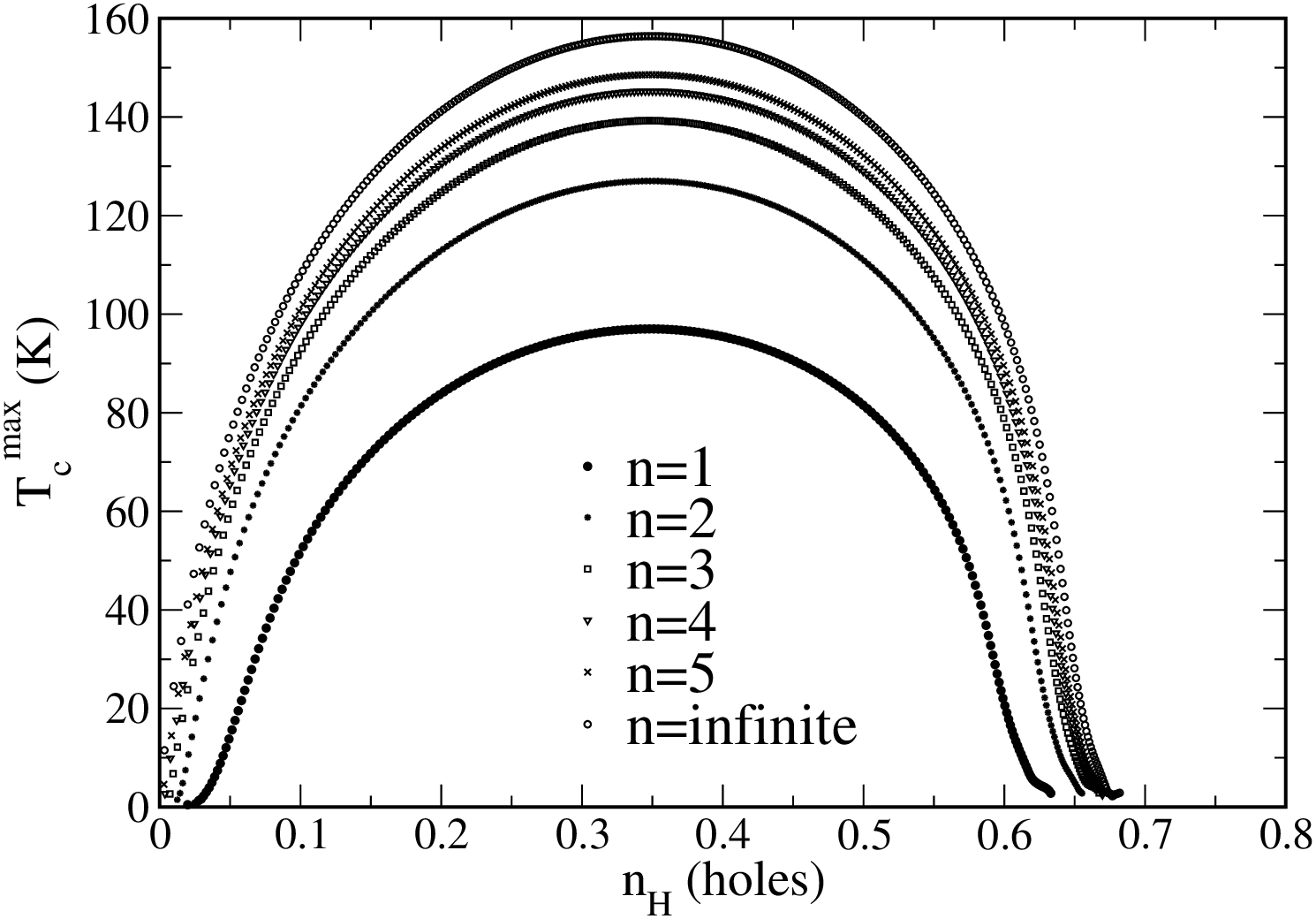}
\end{center}
\caption{ Calculated $T_{c}$ vs $n_{H}$ in HgBa$_{2}$Ca$_{n-1}$Cu$_{n}$O$_{2n+2+\delta}$
as a function of the number of CuO$_{2}$ layers. }
\end{figure}  

The present study shows that interlayer coupling is the driving force for the enhancement of 
$T_{c}^{max}$ for multilayer systems. This does not conflict with the experimental fact that 
$T_{c}^{max}$ saturates as $n \ge 3$. In fact, there exist five-fold (outer) and four-fold 
CuO$_{2}$ (inner) planes surrounded by pyramidal and square oxygens in the multilayer system. 
Investigations carried out by different experimental techniques and model calculations 
\cite{chen1,dist,trok,kote} showed that the distribution of charge carriers are 
nonhomogeneous among the CuO$_{2}$ sheets and the hole concentration in the outer CuO$_{2}$ 
plane is larger than that in the inner CuO$_{2}$ plane. BVS analyses \cite{chen1} and NMR 
studies \cite{kote} on the Hg-based series revealed that the highest $T_{c}^{max}$ 
corresponds to the smallest difference in $n_{H}$ between two types of CuO$_{2}$ planes.
When the number of CuO$_{2}$ layer is larger than three, the reduction of $T_{c}^{max}$ 
comes from the large difference in $n_{H}$ between the outer and inner CuO$_{2}$ planes.
For compounds with more than three CuO$_{2}$ planes, the enhancement of $T_{c}^{max}$ seems 
possible at ambient pressure if one can adequately dope the inner planes.      

In summary, we have investigated the observed $T_{c}$ variation in hole-doped cuprates on 
the basis of an extended interlayer coupling model. We demonstrate that the next nearest 
neighboring hopping $t^{\prime}$ dominates the variation of the maximum $T_{c}$ from series 
to series and the interlayer coupling strength controls the difference of the maximum $T_{c}$ 
among the compounds in a layered homogeneous series. These results provide helpful guidelines 
in the search for new high-$T_{c}$ superconductors. 

The authors are grateful to J. S. Schilling and W. G. Yin for many helpful discussions. This 
work was supported in part by the Earmarked Grant for Research of Project CUHK 4037/02P.

\end{document}